\documentclass[aps,twocolumn,showkeys,superscriptaddress]{revtex4-2}

\usepackage{amssymb}
\usepackage{graphicx}
\usepackage{xspace}
\usepackage{hyperref}
\usepackage{jabbrv}

\newcommand{\xCr}{x_{\rm{Cr}}}
\newcommand{\aCrCr}{{\alpha}_{\rm{CrCr}}}
\newcommand{\Cru}{\rm{Cr}_{\uparrow}}
\newcommand{\Crd}{\rm{Cr}_{\downarrow}}
\newcommand{\aCruu}{\alpha_{\rm{Cr_{\uparrow \uparrow}}}}
\newcommand{\aCrud}{\alpha_{\rm{Cr_{\uparrow \downarrow}}}}
\newcommand{\muB}{\rm{\mu_B}}
\newcommand{\fcc}{\textit{fcc}\xspace}
\newcommand{\hcp}{\textit{hcp}\xspace}

\begin{document}
\title{Magnetically driven short-range order can explain \\ anomalous measurements in CrCoNi}

\author{Flynn Walsh}
\affiliation{Materials Sciences Division, Lawrence Berkeley National Laboratory, Berkeley, CA 94720}
\affiliation{Graduate Group in Applied Science \& Technology, University of California, Berkeley, CA 94720}
\author{Mark Asta}
\affiliation{Materials Sciences Division, Lawrence Berkeley National Laboratory, Berkeley, CA 94720}
\affiliation{Department of Materials Science \& Engineering, University of California, Berkeley, CA 94720}
\author{Robert O. Ritchie}
\email{roritchie@lbl.gov}
\affiliation{Materials Sciences Division, Lawrence Berkeley National Laboratory, Berkeley, CA 94720}
\affiliation{Department of Materials Science \& Engineering, University of California, Berkeley, CA 94720}

\date{March 23, 2021}

\begin{abstract}
The presence, nature, and impact of chemical short-range order in the multi-principal element alloy CrCoNi are all topics of current interest and debate.
First-principles calculations reveal that its origins are fundamentally magnetic, involving repulsion between like-spin Co-Cr and Cr-Cr pairs that is complemented by the formation of a magnetically aligned sublattice of second-nearest-neighbor Cr atoms.
Ordering models following these principles are found to predict otherwise anomalous experimental measurements concerning both magnetization and atomic volumes across a range of compositions.
In addition to demonstrating the impact of magnetic interactions and resulting chemical rearrangement, the possible explanation of experiments would imply that short-range order of this type is far more prevalent than previously realized.
\end{abstract}

\keywords{short-range order; high-entropy alloys; magnetism; frustration}
\maketitle
Multi-principal element alloys (MPEAs), which include high-entropy alloys, have become intensely investigated in recent years as they offer a practically limitless design space that, in the small portion thus far explored, has already yielded several promising materials \cite{yeh2004,miracle2017,george2019,george2020}.
In particular, a large body of research has been devoted to face-centered cubic (\fcc) systems composed of 3$d$ transition metals, namely the equimolar (Cantor) alloy CrMnFeCoNi \cite{cantor2004} and its derivatives.
These ostensibly disordered \fcc MPEAs display highly desirable combinations of mechanical properties that are attributable to deformation mechanisms \cite{george2019,george2020} that can be tuned through careful control of alloy parameters such as chemistry \cite{li2017b} and even magnetic structure \cite{wu2020}. 
Another potentially important, if enigmatic, factor in the engineering of this class of materials is the presence of atomic-scale short-range order (SRO).
In this regard, particular attention has been given to the equiatomic CrCoNi alloy, a representative system that is noteworthy for its cryogenic damage tolerance and general mechanical superiority to the five-component CrMnFeCoNi \cite{wu2014,gludovatz2016,laplanche2017}.

SRO in CrCoNi was theoretically first examined by Tamm et al. \cite{tamm2015} through Monte Carlo (MC) optimization of on-lattice density functional theory (DFT) simulations; similar calculations were later performed by Ding et al. \cite{ding2018}. 
The results of both studies are summarized in Table \ref{table:WC} in terms of nearest-neighbor Warren-Cowley (WC) SRO parameters \cite{cowley1950} (see \textit{Materials and Methods} and Eq. \ref{eq:WC}).
Although extremely limited in statistical sampling, these computations suggest a general trend of additional Cr-Co and Cr-Ni neighbors at the expense of Cr-Cr pairs that is qualitatively supported by analysis of X-ray absorption fine structure \cite{zhang2017}. 
SRO of this form has been predicted to appreciably affect properties ranging from magnetic moment \cite{tamm2015} to stacking fault energy \cite{ding2018}, but the nature of the more than 40 meV per atom driving force (i.e., reduction in energy) observed in MC simulations has not been previously explained.

In the following section, further application of spin-polarized DFT reveals how the dominant bonding preferences of the CrCoNi system originate from magnetic interactions.
Chief among these is the frustration of antiferromagnetic Cr, which can be greatly relieved by the minimization of Cr-Cr nearest neighbors.
Indeed, the frequency of these bonds is shown to fully account for the energies of structures containing previously reported nearest-neighbor ordering.
While Tamm et al. \cite{tamm2015} raised the possibility of magnetic frustration in CrCoNi, the phenomenon has only been directly addressed in a CrCoFeNi alloy, in which the antiferromagnetism of Cr was attributed to the promotion of a chemical Cr L$1_2$ sublattice, as possible given $\xCr = 0.25$ \cite{niu2015}.
Interestingly, CrCoNi is found here to favor a reminiscent sublattice of magnetically aligned Cr atoms beyond the effects of nearest-neighbor interactions. 
Calculations further indicate that magnetically aligned Co-Cr pairs are repulsive, emphasizing the importance of atomic spin polarization in the formation of SRO.

While these observations do not directly address the thermodynamic aspects of order, classical simulations using cluster expansions \cite{pei2020} and the embedded-atom method \cite{li2019a} suggest that significant degrees of SRO may exist even at high temperatures.
Although the specifics of any model omitting explicit magnetic interactions must be interpreted with caution, these approaches should reasonably approximate the relevant energy scales.
Still, despite continuing progress \cite{zhang2020}, direct experimental evidence of SRO in CrCoNi has proven elusive on account of the chemical similarity of its constituent elements.
Recently, Yin et al. \cite{yin2020} have questioned the impact or even existence of SRO in these materials, suggesting that the DFT calculations supporting its existence may be erroneous.
Indeed, computational predictions for random solid solutions of CrCoNi notably contradict experimental measurements of spontaneous magnetization \cite{sales2016,sales2017} and partial molar volumes \cite{yin2020} in ostensibly disordered samples.

Addressing these apparent anomalies, the previously identified ordering principles are applied to nonstoichiometric Cr-Co-Ni compositions in order to replicate the scenarios studied by Refs. \cite{sales2016,yin2020}.
The inclusion of SRO in DFT calculations is shown to theoretically reproduce experimental measurements, offering the possibility that previously examined material contained significant degrees of order.
This interpretation suggests that SRO is not only critical for a wide range of properties, but also is widely prevalent under standard processing conditions.
Several noteworthy implications of these results are then discussed.

\section*{Results}\label{sec:results}
\begin{table}
\caption{\label{table:WC}WC SRO parameters (Eq. \ref{eq:WC}) reported by two previous DFT-MC studies (second and third columns), compared to a simple structural model (fourth column) that is intended to relieve magnetic frustration by eliminating Cr-Cr bonds.}
\begin{ruledtabular}
\begin{tabular}{lccc}
\shortstack{Neighbor \\ Pair \\ {} } & 
\shortstack{Tamm et al. \cite{tamm2015} \\ DFT-MC \\ 500 K }& 
\shortstack{Ding et al. \cite{ding2018} \\ DFT-MC \\ 500 K }& 
\shortstack{Simple \\ structural \\ model} \\
    \hline
    Cr-Cr & 0.42 & 0.40 & $\aCrCr$ \\
    Co-Cr & -0.16 & -0.25 & -$\aCrCr$/2 \\
    Ni-Cr & -0.27 & -0.15 & -$\aCrCr$/2 \\
    Ni-Co & 0.15 & 0.19 & $\aCrCr$/2 \\
    Co-Co & 0.01 & 0.06 & 0.0 \\
    Ni-Ni & 0.12 & -0.04 & 0.0 \\
\end{tabular}
\end{ruledtabular}
\end{table}

\subsection*{Resolving Frustration among Cr Atoms}\label{sec:frustration}
\begin{figure*}
    \centering
    \includegraphics[width=0.75\textwidth]{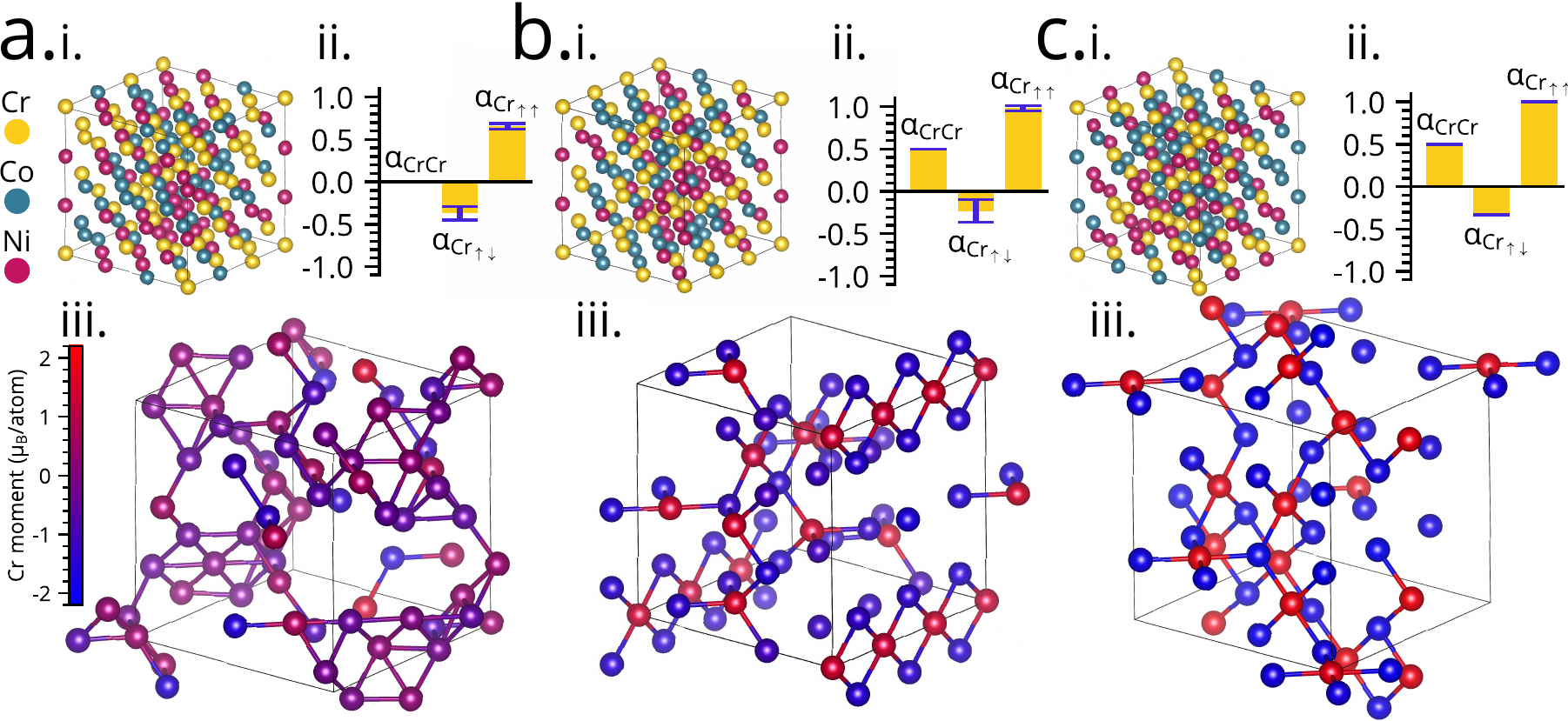}
\caption{\label{fig:frustration}
    An analysis of magnetic furstration in (a) quasirandom, (b) Cr-Cr neighbor-minimized ($\aCrCr = 0.5$ in the simple structural model), and (c) ``spin-ordered" CrCoNi. 
    For each ordering model, (i) an example simulation cell is shown alongside (ii) WC values (Eq. \ref{eq:WC}) for chemical ($\aCrCr$), opposite-spin ($\aCrud$), and same-spin ($\aCruu$) Cr pairs. In (iii), the Cr atoms from (i) are colored by magnetic moment, with nearest-neighbor Cr-Cr bonds drawn.
    Error bars indicate SD from 20 configurations.
    }
\end{figure*}

The magnetic exchange interactions of disordered CrCoNi appear highly frustrated, as exemplified by the quasirandom \cite{zunger1990} configuration depicted in Fig. \ref{fig:frustration}(a)(i).
Fig. \ref{fig:frustration}(a)(iii) shows the local magnetic moments of this cell's Cr atoms, as calculated using spin-polarized DFT, with nearest-neighbor Cr-Cr bonds identified.
Frustration clearly suppresses the local moments of Cr atoms that are bonded to several other Cr nearest neighbors.
In contrast, Cr atoms with fewer Cr neighbors resolve into a network of alternating spins.
Simplistically assigning Cr atoms ``up" and ``down" states from the sign of their local moment enables the calculation of magnetic WC values (Eq. \ref{eq:WC}).
Considering 20 quasirandom configurations, the WC value for same-spin Cr (denoted $\aCruu$) is $0.65 \pm 0.04$, while opposite-spin pairs are commensurately more likely with $\aCrud = -0.37 \pm 0.08$, as graphed in Fig. \ref{fig:frustration}(a)(ii).
Although not accounting for moment magnitudes, these numbers highlight the unfavorability of magnetically aligned Cr pairs.

In a ternary \fcc solid solution of equimolar composition, local chemical ordering can reduce the mean number of same-species nearest neighbors to as low as two ($\aCrCr = 0.5$), offering significant relief from frustration.
In what follows, the effect of Cr neighbor reduction is studied through supercells following a simple structural model in which $\aCrCr$ is the dominant ordering term and other values are nonzero only by conservation of probability (see the fourth column of Table \ref{table:WC} and \textit{Materials and Methods}).

\begin{figure}
    \centering
    \includegraphics[width=\linewidth,height=4in,keepaspectratio]{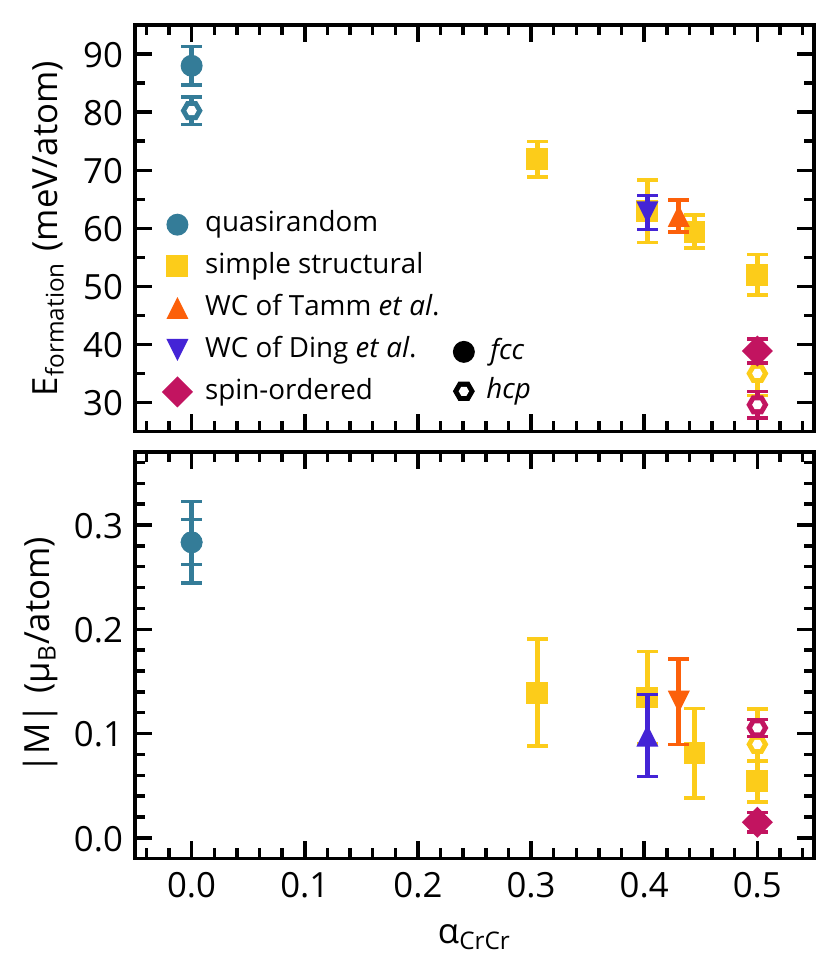}
\caption{\label{fig:energies}
    Formation energy and magnetization for various models of SRO in CrCoNi as a function of the Cr-Cr nearest-neighbor WC SRO parameter (Eq. \ref{eq:WC}). Hollow hexagonal markers represent the model of the corresponding color applied to \hcp lattices. Each datum is the average of 20 configurations; error bars indicate SD.
    }
\end{figure}

Energy and magnetization for configurations with $\aCrCr=0.3,0.4,0.45,0.5$ (plus the quasirandom case of $\aCrCr=0$) are plotted in Fig. \ref{fig:energies}, alongside those of supercells matching the nearest-neighbor WC parameters of Tamm et al. \cite{tamm2015} and Ding et al. \cite{ding2018} reproduced in Table \ref{table:WC}.
It should be emphasized that these are not exact replicas of those studies' configurations; Tamm et al. \cite{tamm2015} report a formation energy of 43.7 meV per atom, substantially lower than the $62.2 \pm 2.7$ meV per atom recalculated presently. 
Nevertheless, the results displayed in Fig. \ref{fig:energies} indicate that, within the margin of error, the energy and magnetization of all these configurations closely follow $\aCrCr$ and that other chemical ordering terms, insofar as they are represented by nearest-neighbor WC parameters, are much less energetically relevant.
In the extreme case of $\aCrCr = 0.5$, formation energy and net moment are $52.0 \pm 3.5$ meV per atom and $ 0.054 \pm 0.04$ $\muB$ per atom, respectively, reduced from $88.0 \pm 3.3$ meV per atom and $0.28 \pm 0.04$ $\muB$ per atom for a quasirandom solution. 

\begin{figure}
    \centering
    \includegraphics[width=\linewidth,height=3in,keepaspectratio]{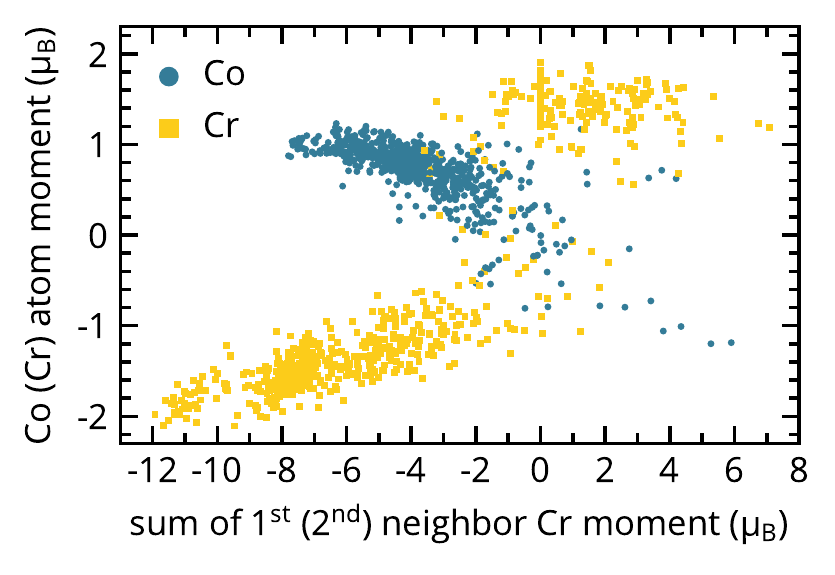}
\caption{\label{fig:moments}
    Atomic moments of Co and Cr atoms from 20 configurations minimizing adjacent Cr in the simple structural model ($\aCrCr = 0.5$).
Co and Cr moments are plotted against the cumulative moments of their nearest-neighbor and second-nearest-neighbor Cr atoms, respectively.
    }
\end{figure}

\subsection*{Accounting for Higher-Order Effects}\label{sec:spin-order}
The discrepancy between the formation energy determined by Tamm et al. \cite{tamm2015} and the value recalculated on the basis of their reported WC parameters suggests the existence of order beyond what can be understood in terms of chemical nearest neighbors.
Furthermore, the range of Co and Cr atomic moments obtained under the simple structural model for $\aCrCr = 0.5$,  presented in Fig. \ref{fig:moments}, indicates that some degree of frustration persists in this regime.
Comparing these values to the moments of neighboring elements reveals trends that offer clues to the physics driving chemical rearrangement.

Specifically, Fig. \ref{fig:moments} shows how Co atoms favor moments antiparallel to those of immediately adjacent Cr.
While Ni atoms possess negligible local moment under all degrees of order, most Co align ferromagnetically, the direction of which will define a reference spin ``up" state, to which Cr atoms are either aligned ($\Cru$) or opposed ($\Crd$).
The preferred antialignment of Cr and Co moments is reflected in Figs. \ref{fig:frustration} and \ref{fig:moments} (as well as Tamm et al. \cite{tamm2015}), where $\Crd$ outnumber $\Cru$ by a factor of three.
Of course, the presence of $\Cru$ is required to minimize the possibility of like-spin Cr pairs; $x_{\Crd} = \frac{1}{4}$ is the maximum possible concentration that can exist on an \fcc lattice without same-species nearest neighbors.
Consequently, the minimum fraction of $\Cru$ is $x_{\Cru} = \xCr - x_{\Crd} = \frac{1}{12}$.

The magnitude of a $\Crd$ moment most strongly depends on none of its nearest neighbors, but rather the magnetization of its Cr second-nearest neighbors, as plotted in Fig. \ref{fig:moments}.
In particular, these data indicate that second-nearest-neighbor sublattices of $\Crd$ (i.e., $\Crd$ with six $\Crd$ second-nearest neighbors) consistently display local moments in the vicinity of -2 $\muB$. 

These preferences motivate a new ``spin-ordered" model that reduces $\Cru$-$\Cru$, $\Crd$-$\Crd$, and Co-$\Cru$ nearest neighbors while maximizing $\Crd$-$\Crd$ second-nearest neighbors (see \textit{Materials and Methods} for a full description).
The average formation energy and net moment for 20 of these supercells are $38.9 \pm 2.0$ meV per atom and $0.015 \pm 0.01$ $\muB$ per atom, respectively.
These values, included in Fig. \ref{fig:energies}, are not only substantially lower than in the simple structural model, but also display minimal spread, implying that the remaining configurational degrees of freedom are not energetically significant.
The specific importance of magnetism is even more explicitly shown in \textit{SI Appendix}, Fig. S1.

Interestingly, both models of SRO appear similarly applicable to hexagonal close-packed (\hcp) lattices, which 0-K DFT predicts to be lower in energy under all degrees of order.
This trend can be seen in Fig. \ref{fig:energies}, which includes the formation energy of quasirandom, $\aCrCr=0.5$, and spin-ordered \hcp configurations at $80.3\pm2.3$, $35.0\pm3.7$, and $29.6\pm2.3$ meV per atom, respectively.
These values are all below those of the corresponding \fcc structures, by a margin ranging from $7.7$ meV per atom in the quasirandom case to $17.0$ meV per atom for the simple structural model.
The net magnetizations of \hcp configurations containing SRO, however, are higher than their \fcc counterparts as the local moments of \hcp $\Crd$ atoms do not realize the same magnitudes, especially in the spin-ordered structures.
In fact, the relatively small energy difference between simple structural and spin-ordered \hcp models suggests that the latter's non-L$1_2$ sublattice lacks the effect of its \fcc counterpart.

\subsection*{Reproducing Magnetization Measurements}\label{sec:magnetization}
\begin{figure}
    \centering
    \includegraphics[width=\linewidth,height=3in,keepaspectratio]{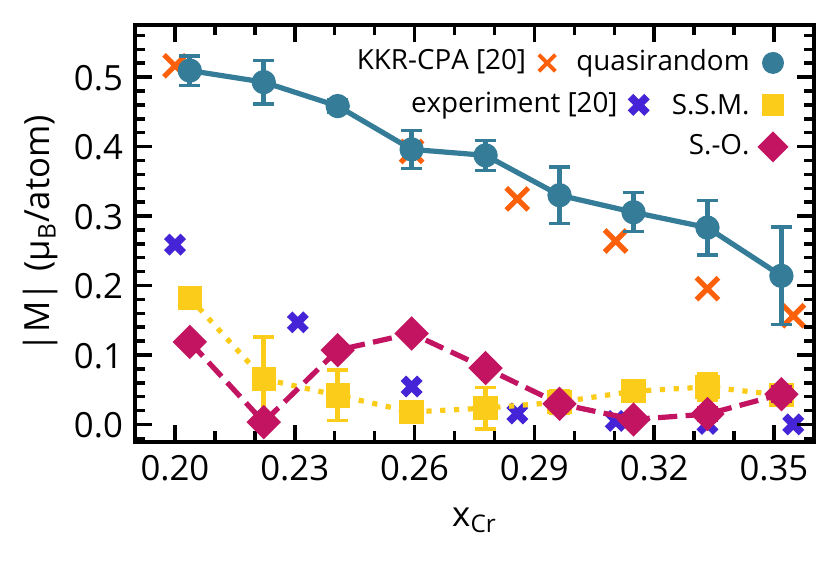}
    \caption{\label{fig:compositions}
    Zero-field and temperature magnetization calculated under several different ordering models for a range of $\xCr$, where $x_{\rm{Co}} = x_{\rm{Ni}} = (1 - \xCr)/2$, compared to 5-K experimental measurements from Ref. \cite{sales2016}.
    "S.S.M." and "S.-O." indicate the ($\aCrCr = 0.5$) simple structural and spin-ordered models, respectively.
    Data points average five configurations, except the equimolar composition, which uses data from Fig. \ref{fig:energies}. Error bars indicate SD.
    }
\end{figure}

Magnetic ordering phenomena have been shown to reduce formation energy, but their realization in real samples of CrCoNi is not well understood.
Indeed, SRO originates from a multitude of thermodynamic and kinetic factors competing throughout a sample's thermal history, which are extremely challenging to model collectively.
Given the comparable level of difficulty faced in the experimental characterization of SRO among these elements, considering indirect evidence for its presence can prove insightful.
In what follows, simulations of structures containing varying degrees of configurational and magnetic order, as described above, provide a potential resolution to discrepancies between magnetization and volume measurements at odds with DFT predictions for random alloys.

One such anomaly was found by Sales et al. \cite{sales2016,sales2017}, in which spontaneous magnetization was measured at 5 K for samples of \fcc Cr$_x$Co$_{(1-x)/2}$Ni$_{(1-x)/2}$ with $0.2 \leq x \leq 0.355$.
Experimental values from Ref. \cite{sales2016} are reproduced in Fig. \ref{fig:compositions}, alongside computational results they obtained using the multiple scattering method of Korring, Kohn, and Rostocker (KKR) with the coherent potential approximation (CPA) \cite{khan2016}, which assumes complete compositional disorder. 
The KKR-CPA results agree well with present calculations for quasirandom configurations, as shown in Fig. \ref{fig:compositions}, but both methods predict magnetizations significantly larger than experimental measurements.

Two further sets of simulation results, each representing a different model of SRO, are included in Fig. \ref{fig:compositions}.
The first dataset (gold squares) applies the simple structural model of maximally eliminating Cr nearest neighbors to additional compositions.
In a similar manner, the second approach (magenta diamonds) uses the chemical and magnetic principles of the spin-ordered state. 
While the spin-ordered model appears closer to experiment around the equimolar composition, the simple structural model's trend better describes the behavior of lower Cr concentrations.
The energy difference between simple structural and spin-ordered models is also notably reduced in this region (\textit{SI Appendix}, Fig. S2).
The possibility that distinct compositions could lead to different degrees of order is crucial for the next set of calculations, in which measurements concerning partial molar volumes are analogously reproduced using SRO.

\subsection*{Reproducing Volume Measurements}
\begin{figure*}
    \centering
    \includegraphics[width=0.75\textwidth]{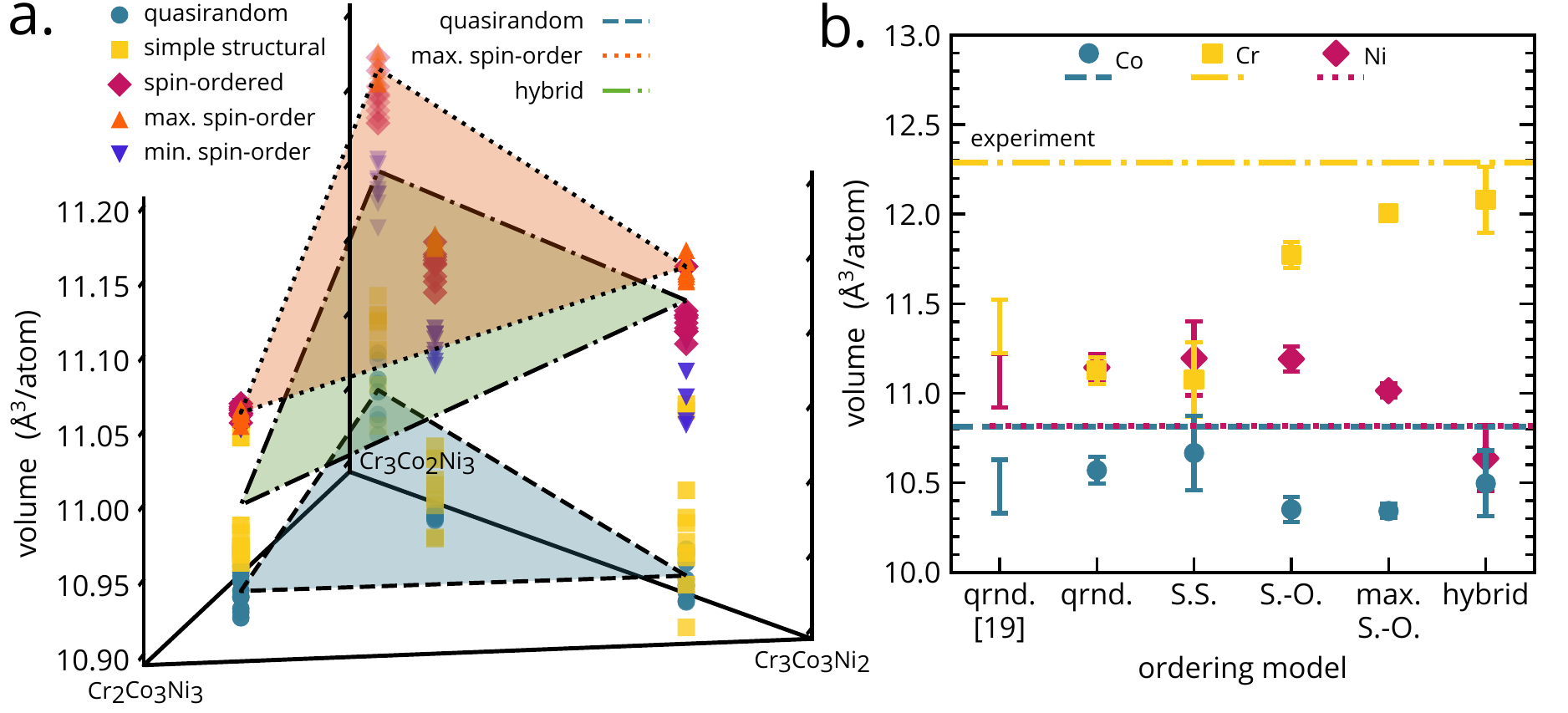}
\caption{\label{fig:volumes}
    (a) Total alloy volume (per atom) for several ordering models applied to CrCoNi, $\rm{Cr_{10} Co_{13} Ni_{13}}$, $\rm{Cr_{13} Co_{10} Ni_{13}}$, and $\rm{Cr_{13} Co_{13} Ni_{10}}$, following Ref. \cite{yin2020}.
    Planes illustrate fits of Eq. \ref{eq:volume} to various models. (b) Partial molar volumes regressed from ordering models shown in (a) or a combination thereof, compared to the experiments of Ref. \cite{yin2020}. ``qrnd." refers to quasirandom configurations, ``S.S.M." stands for the ($\aCrCr = 0.5$) simple structural model, and ``S.-O." denotes spin-ordered. ``min." and ``max." respectively indicate minimum and maximum. 
    The ``hybrid" model considers different ordering models at compositions based on previous results.
    As in Ref. \cite{yin2020}, error bars indicate the 95\% CI.
    }
\end{figure*}

In order to assess lattice misfit and predict yield strength, Yin et al. \cite{yin2020} regressed partial molar volumes of individual chemical species ($V_{\rm{Cr}}$, $V_{\rm{Co}}$, and $V_{\rm{Ni}}$) from the measured total volume ($V_{\rm{alloy}}$) of several samples in close compositional proximity.
Specifically, they determined $V_{\rm{Cr}}$, $V_{\rm{Co}}$, and $V_{\rm{Ni}}$ from four measurements of $V_{\rm{alloy}}$ by fitting to the relation
\begin{equation}\label{eq:volume}
V_{\rm{alloy}} = V_{\rm{Cr}} x_{\rm{Cr}} + V_{\rm{Co}} x_{\rm{Co}} + V_{\rm{Ni}} x_{\rm{Ni}}.
\end{equation} 
Following their approach, structural relaxations were performed for several ordering models applied to CrCoNi, $\rm{Cr_{10} Co_{13} Ni_{13}}$, $\rm{Cr_{13} Co_{10} Ni_{13}}$, $\rm{Cr_{13} Co_{13} Ni_{10}}$.
Resulting $V_{\rm{alloy}}$ values are plotted in the composition space of Fig. \ref{fig:volumes}(a).
In addition to the aforementioned simple structural and spin-ordered models, two variants of the spin-ordered state were also considered, containing either minimal or maximal ordering of Co atoms (see \textit{Materials and Methods} for details).

Several planes representing distinct fits of Eq. \ref{eq:volume} are drawn in Fig. \ref{fig:volumes}(a), including one for quasirandom structures (blue, dashed edges), which were also computationally examined by Yin et al \cite{yin2020}.
For a given ordering model, compositionally disproportionate volume change can tilt the plane of Eq. \ref{eq:volume} relative to the random state, yielding different partial molar volumes. 
While the simple structural model only modestly affects total volume relative to the quasirandom state, the formation of a $\Crd$ sublattice and corresponding growth of those moments (as in Fig. \ref{fig:moments}) leads to significant volumetric dilation, especially at higher Cr concentrations. 
The most extreme shift occurs under maximal spin-ordering, corresponding to the orange plane with a dotted border in Fig. \ref{fig:volumes}.

Fig. \ref{fig:volumes}(b) contains partial molar volumes determined from the different models of order considered in Fig. \ref{fig:volumes}(a).
Results for quasirandom configurations, or even those containing simple structural ($\aCrCr=0.5$) SRO, are not far from the (quasirandom) computational predictions of Ref. \cite{yin2020}.
Volume increases under further degrees of order, however, shift $V_{\rm{Ni}}$ and $V_{\rm{Cr}}$ toward the experimental results of Yin et al. \cite{yin2020}, although some discrepancies remain, particularly for $V_{\rm{Co}}$ \cite{yin2020}.

As demonstrated in the previous subsection, composition-dependent ordering should be considered; while this opens many possibilities, the analysis of magnetization offers some guidance.
Specifically, the data from Fig. \ref{fig:compositions} and \textit{SI Appendix}, Fig. S2 suggest that the spin-ordered model explains the equiatomic point well, while a sample with $\xCr \lesssim 0.3$ is better represented by the simple structural model.
If the configurations with reduced Co similarly contain less ordering of Co (min. spin-order), then a ``hybrid" fit can be constructed, as graphed in Fig. \ref{fig:volumes}(a) (green, dot-dashed border).
More explicitly, the hybrid model assumes simple structural order at $\rm{Cr_{10} Co_{13} Ni_{13}}$, spin-order for CrCoNi and $\rm{Cr_{13} Co_{13} Ni_{10}}$, and ``min. spin-order" at $\rm{Cr_{13} Co_{10} Ni_{13}}$.
Fitting to this combination of models predicts partial molar volumes, shown on the right of Fig. \ref{fig:volumes}(b), that reproduce experiments within established levels of error for this type of calculation \cite{yin2020}.

\section*{Discussion}\label{sec:discussion}
The antiferromagnetic alignment of Cr-Cr and Cr-Co pairs observed in the previous section is expected from Anderson-type theory of magnetic interactions, as is the suppression of Ni moment in the vicinity of these elements \cite{moriya1965}.
However, the central role of magnetism in driving SRO is not fully appreciated.
For instance, our results confirm magnetism as the reason that a recently developed embedded-atom method potential, a purely chemical formalism, predicts a hierarchy of neighbor interactions \cite{li2019a} differing from the DFT-MC studies represented in Table \ref{table:WC}.
The nature of the non-nearest-neighbor interactions motivating a $\Crd$ sublattice is less immediately clear, although its formation is associated with significant changes in the spin-polarized density of states (\textit{SI Appendix}, Fig. S3).
Specifically, the density of states of the positive spin channel approaches zero just above the Fermi level in the spin-ordered model.

It should be emphasized that the proposed ordering principles largely hold in \hcp structures, as shown in Fig. \ref{fig:energies}.
Thus, the positive stacking fault energy determined in Ref. \cite{ding2018} arises simply from the disruption of \fcc SRO by the imposition of \hcp stacking.
As seen in elemental Co and the calculations of Ref. \cite{niu2018}, \fcc CrCoNi stabilizes only at high temperatures, as a consequence of vibrational entropy, suggesting that the phase's low-temperature persistence could be supported by retained ordering.
SRO of this nature could also explain why deforming CrCoNi induces thin layers of \hcp stacking along twin boundaries \cite{miao2017}, where local ordering has been broken, but not any larger regions of the theoretically more stable phase.

In any case, the creation of simple and transferable ordering models can greatly facilitate the study of such scenarios, as demonstrated by the simulation of experiments concerning spontaneous magnetization and partial molar volumes.
The ability of SRO to reproduce these otherwise unexplained measurements offers the possibility that the experimental samples contained order following, to various extents, the previously discussed magnetic principles.
This hypothesis may seem striking given that none of the experimental samples were annealed in a manner intended to promote the development of order.
However, SRO in CrCoNi has a large energetic driving force \cite{tamm2015,ding2018}, requires a very short diffusion length, and is predicted to persist at high temperatures by multiple models \cite{li2019a,pei2020}.
In light of these factors, an extremely rapid onset of SRO during the cooling of samples is not unimaginable. 
Ordering this ubiquitous would be impractical to control and offer little utility for tuning alloy properties, although CrCoNi may be relatively unique in this regard given its high concentration of frustrated Cr.
Of course, further experimental work is needed to confidently characterize the material's chemical and magnetic structure.

Sales et al. \cite{sales2017} examine their samples with high-resolution electron microscopy, ruling out appreciable local clustering or compositional heterogeneity, but not necessarily atomic-scale chemical SRO. 
Yin et al. \cite{yin2020} argue that their measurements will be unaffected given the minimal volume change associated with the SRO of Ding et al. \cite{ding2018}.
However, the DFT-MC configurations of Ref. \cite{ding2018} are likely not fully equilibrated due to the small number of DFT-MC steps achievable and do not include the spin ordering responsible for the greatest volume changes in the present calculations.

It is also worth stating that the ordering models studied in this investigation aim to approximate the realization of energetically favorable trends, not confidently assert the condition of specific samples.
For example, while the spin-ordered state maximizes $\Crd$ second-nearest neighbors with a sublattice, a continuous, long-range ordered state is not necessarily expected.
Regardless of how well any individual sample is represented, the potential impact of SRO on a wide-range of properties has been demonstrated.
In contrast, complete compositional disorder is routinely assumed in the modeling of high-entropy alloys, as seen in the referenced examples and countless others.
At the very least, this study indicates that randomness should not be supposed without careful experimental characterization of atomic-scale structure.

Beyond the present analysis, the prospect of magnetically driven SRO raises a number of interesting questions that follow an emerging trend of bridging the conventionally disparate domains of structural and magnetic properties \cite{niu2015,niu2018,wu2020}.
For instance, the energy cost associated with Cr agglomeration should hinder the formation of any lower energy Cr-rich phase via a martensitic mechanism, perhaps buoying the alloy's metastability.
The bulk energy of CrCoNi has been shown to closely follow the frequency of Cr-Cr neighbors---can the same connection be made for its stacking fault energies? 
As a potential corollary, will Cr spins reorient as their local environment is changed by the transmission of dislocations---and how will the dynamics of this process affect slip properties?
Finally, given the more prevalent role of magnetism at low temperatures, could it be connected to the material's superior mechanical performance under cryogenic conditions?

\section*{Conclusion}
In CrCoNi, SRO forms to accommodate the magnetic preferences of its constituent elements, which can be understood as follows.
Co atoms align ferromagnetically, while Ni moment is suppressed.
Individual Cr atoms would prefer to magnetize opposite the direction of Co moments, but the formation of like-spin Cr bonds is extremely unfavorable.
A random \fcc solid solution of these elements therefore exhibits frustration that can be greatly relieved by the minimization of Cr-Cr nearest neighbors, which is shown to be the dominant ordering trend.
Same-spin Co-Cr pairs can be further reduced by swapping specific Co atoms with Ni.
Lastly, at higher Cr concentrations, the majority-spin Cr prefer to maximize their second-nearest-neighbor alignment, which appreciably increases local moment and volume.

This investigation highlights not only the dominant role of magnetism in the SRO of a widely studied MPEA, but also the need to analyze the ordering of similar materials with awareness of atoms' spin polarization in addition to chemical species.
Application of ordering is found to possibly explain discrepancies between two sets of experimental measurements and previous DFT calculations assuming random substitutional disorder.
These results imply that magnetically driven SRO could be widely prevalent in CrCoNi and related systems.

\section*{Materials and Methods}
WC SRO parameters measure the frequency of a chemical pairing relative to that expected in a random solution. 
For adjacent species $i$ and $j$, the latter with concentration $c_j$, the parameter is expressed as
\begin{equation}\label{eq:WC}
    \alpha_{ij} = 1 - \frac{P(j \mid i)}{c_j} 
                = 1 - \frac{P(j \cap i)}{P(j)P(i)},
\end{equation}
where $P(j \mid i)$ represents the probability of finding species $j$ adjacent to species $i$ and $P(j \cap i$) is the fraction of $ij$-type nearest-neighbor bonds.
Negative values indicate more neighboring $i$ and $j$ than in a random alloy (corresponding to $\alpha_{ij} = 0$ for all neighbor types) and positive the inverse.
In Fig. \ref{fig:frustration}, $\aCruu$ and $\aCrud$ are calculated as a weighted average of the relevant quantity for each spin.
Tamm et al. \cite{tamm2015} report WC parameters for several temperatures, but given these simulations' small number of steps, the most ordered values (500 K) were selected for Table \ref{table:WC}.
SRO configurations were generated by targeting desired WC values through simulated annealing.
Quasirandom order means $\alpha_{ij} = 0$ for all chemical species $i$ and $j$.
The simple structural model is singularly parameterized by $\aCrCr$; other values follow per Table \ref{table:WC}, although the exact form of these probability-conserving terms appears relatively unimportant.
In the spin-ordered model, the alignment of second-nearest-neighbor $\Crd$ is maximized such that they form a sublattice (L1$_2$ for \fcc), as depicted in Fig. \ref{fig:frustration}(c).
These $\Crd$ represent $\frac{3}{4}$ of all Cr, while the remaining $\frac{1}{4}$ become $\Cru$ that are randomly distributed while avoiding nearest neighbors.
Ni and Co are assigned to remaining sites to minimize unfavorable Co-$\Cru$ pairs.
The ``min. spin-order" configurations included in Fig. \ref{fig:volumes} distribute Co atoms to available sites randomly, while the ``max. spin-order" model optimizes the placements of Co and $\Cru$ simultaneously, fully segregating these species.
All \fcc configurations contained 108 atoms (a $3\times3\times3$ conventional unit cell), while \hcp cells had 92. Twenty cells were used per datum in Fig. \ref{fig:energies}, while five were used for each ordering model and composition in Figs. \ref{fig:compositions} \& \ref{fig:volumes}, with the exception of the equimolar composition in Fig. \ref{fig:compositions}, which uses data from Fig. \ref{fig:energies}.
Collinearly spin-polarized DFT structure optimizations were performed \cite{towns2014} using the Vienna Ab initio Simulation Package (VASP) \cite{kresse1993,kresse1996,kresse1996a} with Perdew, Burke, and Ernzerhof's parametrization of the generalized gradient approximation \cite{perdew1996} and projector-augmented wave potentials \cite{kresse1999}.
Electronic states were sampled in reciprocal space with a $3\times3\times3$ Monkhorst-Pack grid and 420-eV plane wave cutoff.
Calculations are further discussed in \textit{SI Appendix, SI Text}.

\subsection*{Data Availability}
VASP data have been deposited in MPContribs (\url{https://contribs.materialsproject.org/projects/mpea_sro/}) \cite{walsh2021}.

\begin{acknowledgments}
This work was supported by the U.S. Department of Energy, Office of Basic Energy Sciences, Materials Sciences and Engineering Division, under contract No. DE-AC02-05CH11231 within the Damage-Tolerance in Structural Materials (KC13) program.
Simulations used resources provided by the National Energy Research Scientific Computing Center, a US Department of Energy Office of Science User Facility operated under the same contract number.
Calculations were also performed at the Texas Advanced Computing Center at the University of Texas at Austin, as part of Extreme Science and Engineering Discovery Environment, which is supported by NSF Grant ACI-1548562.
F.W. was supported by the Department of Defense through the National Defense Science \& Engineering Graduate Fellowship Program and credits Anirudh R. Natarajan and Anton van der Ven for spurring his interest in the magnetic interactions of this system.
\end{acknowledgments}

\bibliographystyle{style}
\bibliography{main}

\begin{thebibliography}{10}

\bibitem{yeh2004}
J.-W. Yeh, et~al., Nanostructured {{High}}-{{Entropy Alloys}} with {{Multiple
  Principal Elements}}: {{Novel Alloy Design Concepts}} and {{Outcomes}}.
\newblock {\em\protect\JournalTitle{Advanced Engineering Materials}}
  \textbf{6}, 299--303 (2004).

\bibitem{miracle2017}
D.~Miracle, O.~Senkov, A critical review of high entropy alloys and related
  concepts.
\newblock {\em\protect\JournalTitle{Acta Materialia}} \textbf{122}, 448--511
  (2017).

\bibitem{george2019}
E.~P. George, D.~Raabe, R.~O. Ritchie, High-entropy alloys.
\newblock {\em\protect\JournalTitle{Nature Reviews Materials}} \textbf{4},
  515--534 (2019).

\bibitem{george2020}
E.~George, W.~Curtin, C.~Tasan, High entropy alloys: {{A}} focused review of
  mechanical properties and deformation mechanisms.
\newblock {\em\protect\JournalTitle{Acta Materialia}} \textbf{188}, 435--474
  (2020).

\bibitem{cantor2004}
B.~Cantor, I.~Chang, P.~Knight, A.~Vincent, Microstructural development in
  equiatomic multicomponent alloys.
\newblock {\em\protect\JournalTitle{Materials Science and Engineering: A}}
  \textbf{375-377}, 213--218 (2004).

\bibitem{li2017b}
Z.~Li, F.~K{\"o}rmann, B.~Grabowski, J.~Neugebauer, D.~Raabe, Ab initio
  assisted design of quinary dual-phase high-entropy alloys with
  transformation-induced plasticity.
\newblock {\em\protect\JournalTitle{Acta Materialia}} \textbf{136}, 262--270
  (2017).

\bibitem{wu2020}
X.~Wu, et~al., Role of magnetic ordering for the design of quinary
  {{TWIP}}-{{TRIP}} high entropy alloys.
\newblock {\em\protect\JournalTitle{Physical Review Materials}} \textbf{4},
  033601 (2020).

\bibitem{wu2014}
Z.~Wu, H.~Bei, F.~Otto, G.~Pharr, E.~George, Recovery, recrystallization, grain
  growth and phase stability of a family of {{FCC}}-structured multi-component
  equiatomic solid solution alloys.
\newblock {\em\protect\JournalTitle{Intermetallics}} \textbf{46}, 131--140
  (2014).

\bibitem{gludovatz2016}
B.~Gludovatz, et~al., Exceptional damage-tolerance of a medium-entropy alloy
  {{CrCoNi}} at cryogenic temperatures.
\newblock {\em\protect\JournalTitle{Nature Communications}} \textbf{7}, 10602
  (2016).

\bibitem{laplanche2017}
G.~Laplanche, et~al., Reasons for the superior mechanical properties of
  medium-entropy {{CrCoNi}} compared to high-entropy {{CrMnFeCoNi}}.
\newblock {\em\protect\JournalTitle{Acta Materialia}} \textbf{128}, 292--303
  (2017).

\bibitem{tamm2015}
A.~Tamm, A.~Aabloo, M.~Klintenberg, M.~Stocks, A.~Caro, Atomic-scale properties
  of {{Ni}}-based {{FCC}} ternary, and quaternary alloys.
\newblock {\em\protect\JournalTitle{Acta Materialia}} \textbf{99}, 307--312
  (2015).

\bibitem{ding2018}
J.~Ding, Q.~Yu, M.~Asta, R.~O. Ritchie, Tunable stacking fault energies by
  tailoring local chemical order in {{CrCoNi}} medium-entropy alloys.
\newblock {\em\protect\JournalTitle{Proceedings of the National Academy of
  Sciences}} \textbf{115}, 8919--8924 (2018).

\bibitem{cowley1950}
J.~M.~F. Cowley, An approximate theory of order in alloys.
\newblock {\em\protect\JournalTitle{Physical Review}} \textbf{77}, 669 (1950).

\bibitem{zhang2017}
F.~X. Zhang, et~al., Local {{Structure}} and {{Short}}-{{Range Order}} in a
  {{NiCoCr Solid Solution Alloy}}.
\newblock {\em\protect\JournalTitle{Physical Review Letters}} \textbf{118}
  (2017).

\bibitem{niu2015}
C.~Niu, et~al., Spin-driven ordering of {{Cr}} in the equiatomic high entropy
  alloy {{NiFeCrCo}}.
\newblock {\em\protect\JournalTitle{Applied Physics Letters}} \textbf{106},
  161906 (2015).

\bibitem{pei2020}
Z.~Pei, R.~Li, M.~C. Gao, G.~M. Stocks, Statistics of the {{NiCoCr}}
  medium-entropy alloy: {{Novel}} aspects of an old puzzle.
\newblock {\em\protect\JournalTitle{npj Computational Materials}} \textbf{6},
  1--6 (2020).

\bibitem{li2019a}
Q.-J. Li, H.~Sheng, E.~Ma, Strengthening in multi-principal element alloys with
  local-chemical-order roughened dislocation pathways.
\newblock {\em\protect\JournalTitle{Nature Communications}} \textbf{10}, 1--11
  (2019).

\bibitem{zhang2020}
R.~Zhang, et~al., Short-range order and its impact on the {{CrCoNi}}
  medium-entropy alloy.
\newblock {\em\protect\JournalTitle{Nature}} \textbf{581}, 283--287 (2020).

\bibitem{yin2020}
B.~Yin, S.~Yoshida, N.~Tsuji, W.~A. Curtin, Yield strength and misfit volumes
  of {{NiCoCr}} and implications for short-range-order.
\newblock {\em\protect\JournalTitle{Nature Communications}} \textbf{11}, 2507
  (2020).

\bibitem{sales2016}
B.~C. Sales, et~al., Quantum {{Critical Behavior}} in a {{Concentrated Ternary
  Solid Solution}}.
\newblock {\em\protect\JournalTitle{Scientific Reports}} \textbf{6} (2016).

\bibitem{sales2017}
B.~C. Sales, et~al., Quantum critical behavior in the asymptotic limit of high
  disorder in the medium entropy alloy {{NiCoCr0}}.8.
\newblock {\em\protect\JournalTitle{npj Quantum Materials}} \textbf{2}, 33
  (2017).

\bibitem{zunger1990}
A.~Zunger, S.-H. Wei, L.~G. Ferreira, J.~E. Bernard, Special quasirandom
  structures.
\newblock {\em\protect\JournalTitle{Physical Review Letters}} \textbf{65},
  353--356 (1990).

\bibitem{khan2016}
S.~N. Khan, J.~B. Staunton, G.~M. Stocks, Statistical physics of multicomponent
  alloys using {{KKR}}-{{CPA}}.
\newblock {\em\protect\JournalTitle{Physical Review B}} \textbf{93}, 054206
  (2016).

\bibitem{moriya1965}
T.~Moriya, Ferro- and {{Antiferromagnetism}} of {{Transition Metals}} and
  {{Alloys}}.
\newblock {\em\protect\JournalTitle{Progress of Theoretical Physics}}
  \textbf{33}, 157--183 (1965).

\bibitem{niu2018}
C.~Niu, C.~R. LaRosa, J.~Miao, M.~J. Mills, M.~Ghazisaeidi, Magnetically-driven
  phase transformation strengthening in high entropy alloys.
\newblock {\em\protect\JournalTitle{Nature Communications}} \textbf{9}, 1363
  (2018).

\bibitem{miao2017}
J.~Miao, et~al., The evolution of the deformation substructure in a
  {{Ni}}-{{Co}}-{{Cr}} equiatomic solid solution alloy.
\newblock {\em\protect\JournalTitle{Acta Materialia}} \textbf{132}, 35--48
  (2017).

\bibitem{towns2014}
J.~Towns, et~al., {{XSEDE}}: {{Accelerating Scientific Discovery}}.
\newblock {\em\protect\JournalTitle{Computing in Science Engineering}}
  \textbf{16}, 62--74 (2014).

\bibitem{kresse1993}
G.~Kresse, J.~Hafner, Ab initio molecular dynamics for liquid metals.
\newblock {\em\protect\JournalTitle{Physical Review B}} \textbf{47}, 558--561
  (1993).

\bibitem{kresse1996}
G.~Kresse, J.~Furthm{\"u}ller, Efficiency of ab-initio total energy
  calculations for metals and semiconductors using a plane-wave basis set.
\newblock {\em\protect\JournalTitle{Computational Materials Science}}
  \textbf{6}, 15--50 (1996).

\bibitem{kresse1996a}
G.~Kresse, J.~Furthm{\"u}ller, Efficient iterative schemes for ab initio
  total-energy calculations using a plane-wave basis set.
\newblock {\em\protect\JournalTitle{Physical Review B}} \textbf{54},
  11169--11186 (1996).

\bibitem{perdew1996}
J.~P. Perdew, K.~Burke, M.~Ernzerhof, Generalized {{Gradient Approximation Made
  Simple}}.
\newblock {\em\protect\JournalTitle{Physical Review Letters}} \textbf{77},
  3865--3868 (1996).

\bibitem{kresse1999}
G.~Kresse, D.~Joubert, From ultrasoft pseudopotentials to the projector
  augmented-wave method.
\newblock {\em\protect\JournalTitle{Physical Review B}} \textbf{59}, 1758--1775
  (1999).

\bibitem{walsh2021}
F.~Walsh, M.~Asta, R.~O. Ritchie, Short-range order in multi-principal element
  alloys (https://contribs.materialsproject.org/projects/mpea\_sro/) (2021).

\end{thebibliography}

\end{document}